\documentclass[epj]{svjour}
\usepackage{amsmath,graphics,epsfig}
\newcommand{\beqy}{\begin{eqnarray}}
\newcommand{\eeqy}{\end{eqnarray}}
\newcommand{\bmlet}{\begin{subequations}}
\newcommand{\emlet}{\end{subequations}}

\begin{document}
\title{Symmetry energy: nuclear masses and neutron stars}
\author{J.~M. Pearson\inst{1} \and N. Chamel\inst{2} \and A.~F. Fantina\inst{2} \and S. Goriely\inst{2}
}                     
%
%
\institute{D\'ept. de Physique, Universit\'e de Montr\'eal, Montr\'eal 
(Qu\'ebec), H3C 3J7 Canada \and Institut d'Astronomie et d'Astrophysique, CP-226,
Universit\'e Libre de Bruxelles, 1050 Brussels, Belgium}
\date{Received: date / Revised version: date}
%
\abstract{We describe the main features of our most recent Hartree-Fock-Bogoliubov 
nuclear mass models, based on 16-parameter generalized Skyrme forces. They
have been fitted to the data of the 2012 Atomic Mass Evaluation, and favour a
value of 30 MeV for the symmetry coefficient $J$, the corresponding 
root-mean square deviation being 0.549 MeV. We find that this 
conclusion is compatible with measurements of neutron-skin thickness. By 
constraining the underlying interactions to fit various equations of state of 
neutron matter calculated {\it ab initio} our models are well adapted to a realistic
and unified treatment of all regions of neutron stars. We use our models to 
calculate the composition, the equation of state, the mass-radius relation and the
maximum mass. Comparison with observations of neutron stars again favours a value
of $J$ = 30 MeV.}


\PACS{
      {21.65.Ef}{Symmetry energy}   \and
      {21.10.Dr}{Binding energies and masses} \and
      {97.60.Jd}{Neutron stars}   \and
      {21.60.Jz}{Nuclear Density Functional Theory and extensions}   \and
      {26.60.Kp}{Equations of state of neutron-star matter}
     } 

\maketitle

\section{Introduction}
\label{intro}
The concept of nuclear symmetry energy probably arises most naturally in 
connection with the liquid-drop model of the nucleus, which was inspired in 
large part by the observation that the sharp radius $R$ of any nucleus with $Z$
protons and $N$ neutrons is given by $R \simeq r_0A^{1/3}$, where the mass
number $A = N + Z$ and $r_0$ is a constant. The roughly constant density is 
what one would expect in view of the saturated character of nuclear forces, 
i.e., their finite range and strong short-range repulsion, and suggests in the
first instance that nuclei consist simply of differently sized pieces of a 
homogeneous substance, ``nuclear matter". In this extreme picture the internal
energy $e$ per nucleon would be the same for all nuclei, but the liquid-drip 
model takes into account surface tension, the non-saturating character of the 
Coulomb force and a ``composition" dependence, writing 
\beqy\label{1}
e = a_v +JI^2 + \left(a_{sf} + a_{ss}I^2\right)A^{-1/3} + 
\frac{3e^2}{5r_0}Z^2A^{-4/3} \, .
\eeqy
With $I = (N - Z)/A$, this expression admits so-called volume-symmetry and
surface-symmetry terms; the dominant contribution to both the kinetic-energy
and potential-energy parts of these terms comes from the Pauli principle,
although the spin dependence and finite range of the nuclear forces play a 
role also. Fitting 
the five parameters of Eq.~(\ref{1}) to the data of the 2012 Atomic Mass 
Evaluation (AME)~\cite{ame12}, we find: $a_v$ = -15.65 MeV,
$J$ = 27.30 MeV, $a_{sf}$ = 17.58 MeV, $a_{ss}$ = - 24.16 MeV and 
$r_0$ = 1.230 fm; the root-mean square (rms) deviation of this fit to 2353 nuclei 
is 3.03 MeV. 

It is the sum of the first two terms of the right-hand side of Eq.~(\ref{1}), 
$a_v +JI^2$, that represents, in the absence of any Coulomb effects, the energy
per nucleon of homogeneous or infinite nuclear matter, the density of which is 
$n_0 = 3/(4\pi r_0^{3})$, i.e.,
around 0.16 nucleons per cubic fermis. However, if we let $A$ become very big 
while keeping $I$, and hence $Z/A$, constant Eq.~(\ref{1}) goes over into
\beqy\label{2}
e = a_v +JI^2 + \frac{3e^2}{20r_0}(1 - I)^2A^{2/3} \quad ,
\eeqy
which means that the energy per nucleon diverges in this limit, because of the 
Coulomb term. In a purely neutronic system, $I$ = 1, the Coulomb term vanishes,
but then in view of the above empirical values of the drop-model parameters the
energy per nucleon, $e = a_v +J$, is positive, i.e., the system is unbound. 
Thus it appears that because of the joint action of the Coulomb term and 
volume-symmetry term it is impossible to have a piece of nuclear matter large 
enough for all but the volume terms in the energy to be negligible. 

However, let us put gravity into the drop-model expression 
(\ref{2})~\cite{pea01}. Since this, although attractive, has the same 
non-saturating character as the Coulomb force, Eq.~(\ref{2}) is replaced by
\beqy\label{3}
e = a_v +JI^2 + \frac{3}{5r_0}
\left\{\frac{e^2}{4}(1 - I)^2 - GM^2\right\}A^{2/3} \quad ,
\eeqy
where $G$ is the gravitational constant and $M$ the nucleon mass.
As long as protons are present, i.e., $I < 1$, the Coulomb term will totally
overwhelm the gravitational term and the system will be unbound. But using the
above values of the parameters we find that a purely neutronic system will be
gravitationally bound for $A>3.4\times 10^{56}$, i.e., for a mass in excess of 
$5.6\times 10^{29}$ kg, which is within an order of magnitude of the mass of a 
typical neutron star. Furthermore, the nuclear density $n_0$ is within an order
of magnitude that of neutron stars, typical radii being around 10 km.

This picture of a neutron star as a single giant nucleus, bound not by nuclear 
forces but by gravity is, of course, grossly oversimplified. In the first 
place, the star does not consist entirely of neutrons, but rather of a neutral 
mixture of neutrons, protons and leptons (electrons and muons) in beta equilibrium. 
Secondly, while the bulk of the star is 
indeed homogeneous the density is by no means constant, falling steadily from 
several times the nuclear density $n_0$ at the centre to zero at the surface. 
Moreover, the equation of hydrostatic equilibrium that the density variations 
must satisfy can only be formulated within the framework of general relativity,
one consequence of which is that there is an upper limit on the gravitational 
mass of neutron stars. (However, the fact that there is also a {\it lower}
limit can be understood qualitatively by reference to Eq.~(\ref{3}): if the 
mass of the star falls below a certain value it is blown apart by the symmetry
energy.) 

Actually, when the density falls below about $n_0/2$ 
the homogeneous medium breaks up into inhomogeneities. This marks the
transition from the core to the crust of the star (see, e.g., Ref.~\cite{lrr} 
for a review). The inner part of the crust consists of neutron-proton clusters
along with free neutrons and electrons, while the outer part, below a density 
of about $2.6 \times 10^{-4}$ nucleons per cubic fermis, consists of 
just bound nuclei and electrons, with no free neutrons remaining. The nuclei 
found in this region will generally be highly neutron rich, and those at the 
interface will be on the neutron drip line. As the surface is approached the nuclei 
become less and less neutron rich, and at the surface of the star only nuclei that 
are stable under natural terrestrial conditions are found.

However, all but a tiny fraction of the nucleons of a neutron star are located 
in its core, which means that global properties such as the maximum possible 
mass are determined essentially by the 
properties of homogeneous beta-equilibrated nuclear matter, which we refer to 
as neutron-star matter. To discuss these latter properties it is convenient to
begin by neglecting the presence of electrons and muons, even though they play
a crucial role (it is trivially easy to take account of them), and simply 
generalize to arbitrary density the notion of the hypothetical
Coulomb-free infinite nuclear matter (INM) introduced in connection with the
first two terms of the liquid-drop expression~(\ref{1}). For INM of proton 
density $n_p$ and neutron density $n_n$ we write the equation of state (EoS)
i.e., the energy per nucleon as a function of the total density 
$n = n_p + n_n$ and charge asymmetry $\eta = (n_n - n_p)/n$ (throughout this paper
we assume zero temperature), in the form
\beqy\label{4}
e(n,\eta)=e(n,\eta=0)+ S_1(n)\eta^2+ O\Big(\eta^4\Big) \quad ,
\eeqy
in which the first term on the right-hand side is just the energy per nucleon
of charge-symmetric INM. Identifying $n_0$ as the equilibrium density of 
charge-symmetric INM, we then expand $e(n,\eta=0)$ and the symmetry energy
$S_1(n)$ about $n_0$ in powers of $\epsilon = (n - n_0)/n_0$, thus
\bmlet
\beqy \label{5a}
e(n,\eta=0) = a_v + \frac{1}{18}K_v\epsilon^2 -
\frac{1}{162}K^\prime\,\epsilon^3 + ...
\eeqy
and
\beqy \label{5b}
S_1(n) = J + \frac{1}{3}L\epsilon + \frac{1}{18}K_{sym}\epsilon^2
+ ... \quad .
\eeqy
\emlet
Of particular interest in the context of neutron stars is the special case of
$\eta = 1$, i.e., pure neutron matter (NeuM), the energy per neutron of which
we write as 
\beqy\label{6}
e(n,\eta = 1) = e(n,\eta = 0) + S_2(n)\quad .
\eeqy
Both $S_1(n)$ and $S_2(n)$ have been generally referred to as the 
``symmetry energy", but, because of the terms $O\Big(\eta^4\Big)$ in 
Eq.~(\ref{4}), these two functions are not identical. 
Actually, at nuclear densities the difference is very small but it can become
significant at higher densities (see, for example, Section III of 
Ref.~\cite{gcp10}); in any case, it would be more appropriate to speak of
``asymmetry energy". 

A crucial quantity in discussing the internal hydrostatic equilibrium in 
neutron stars is the pressure, given in homogeneous INM by 
$P = n^2\left\{\partial\,e(n,\eta)/\partial\,n\right\}_\eta$ (in actual neutron-star
matter the
lepton pressure has to be added to this). For $n$ close to
$n_0$ in NeuM this reduces to $P = Ln_0/3$, whence the
importance that has been attached to the $L$ parameter. However, in the more
general case the pressure depends not only on the symmetry energy (of either 
kind) but also on the energy per nucleon of charge-symmetric INM, 
$e(n,\eta=0)$. Thus we shall be concerned here not just with symmetry energy
but with charge-asymmetric INM in general.

Laboratory experiments provide us with considerable, but by no means complete, 
information on the EoS of INM up to a density of around $4$ or $5 n_0$ 
(see, e.g., Ref.~\cite{tsang2012} for a recent review of different methods, 
and Section II of this volume). In this paper it is nuclear masses that 
provide the principal source of laboratory data on which we will draw, 
although we will also examine the implications of neutron-skin measurements.

But in the heavier neutron stars the core density can reach $10 n_0$,
and information on the EoS at such high densities can only be obtained through 
{\it ab initio} many-body calculations of INM starting from realistic 
nucleonic interactions,
as determined by nucleon-nucleon scattering and the bound states of two- and
three-nucleon systems, with possibly some guidance from meson-exchange theory
or QCD (see, e.g., Section I of this volume). In principle, these calculations 
should be 
performed as a function of both density $n$ and asymmetry $\eta$, although it 
should be noted that the calculations are more reliable for the case 
$\eta = 1$, i.e., pure NeuM, for the following reasons: i) two-nucleon 
scattering determines the
phase parameters better in the $T$ = 1 than in the $T$ = 0 states; ii) the
strongly tensor-coupled $^3$S$_1$ and $^3$D$_1$ states are absent; iii) the
relatively poorly determined three-nucleon force is less important in this case
(it would not contribute at all to NeuM if it had zero range).

Besides homogeneous INM, {\it ab initio} calculations of this kind 
have been carried out only for light nuclei and a few medium-mass nuclei, 
and thus cannot be used for either the inner or outer crusts. 
However, for the EoS of the outer crust all that is
required are the masses of the appropriate nuclei, and for the outermost layers
these are known experimentally, namely~\cite{rhs06,pgc11}: $^{56}$Fe, 
$^{62}$Ni, $^{64}$Ni, $^{66}$Ni, $^{86}$Kr, $^{84}$Se, $^{82}$Ge and 
$^{80}$Zn. For the more neutron-rich nuclei of the deeper layers of 
the outer crust recourse must be made to theoretical mass tables, whose 
extrapolations from the data may be expected to be more reliable the more they are
microscopically based. The most 
microscopic atomic mass models currently available are based on the 
Hartree-Fock-Bogoliubov (HFB) method with effective interactions of the Skyrme
or Gogny type. The rms deviations of the most accurate models lie below 
0.6~MeV~\cite{gcp10}. As mass measurements are pushed ever closer to the neutron 
drip line we can expect that at some time in the future the masses of all the nuclei
found in the outer crust will have been measured, making the use of theoretical mass
models in this region unnecessary~\cite{wolf13}. 

On the other hand, the inhomogeneous nuclear matter found in the inner crust 
of a neutron star cannot be reproduced in the laboratory, and the resort to theory
is inevitable. Since the seminal work of Negele and 
Vautherin~\cite{nv73} mean-field calculations have been generally performed 
within the Wigner-Seitz framework~\cite{bal07,gms2011}: 
the crust is replaced by a set of independent spherical cells each of which 
contains only one cluster. However this approach can only be reliably applied 
in the shallowest layers of the inner crust because of the appearance of 
spurious neutron quantum shell effects~\cite{cha07}. These effects persist in 
calculations using a cubic cell with periodic boundary 
conditions~\cite{mag02,gog08,new09}, and can only be properly eliminated by 
using the band theory of solids~\cite{cha07}. However, the calculations would 
then become computationally prohibitive. In previous papers on the inner
crust~\cite{ons08,pcgd12} we have adopted the simpler approach of neglecting 
entirely neutron shell effects, and applying the semi-classical extended 
Thomas-Fermi method. Proton shell effects, which, unlike neutron shell 
effects, are generally not negligible, are easily taken into account via the 
Strutinsky integral theorem. This method is extended in the present paper to 
include pairing correlations, treated within the BCS approximation as discussed in 
Ref.~\cite{pea91}. While
still limited to spherical symmetry this restriction is not
expected to have a significant impact on the calculated EoS. 
 
Effective interactions used in the outer and inner parts of the crust can also
provide a convenient parametrization of the EoS of INM, as obtained from 
realistic calculations. In this way, it is possible to construct unified EoSs 
of neutron stars ensuring a thermodynamically consistent treatment of the 
transitions between the different regions of the star. Such a treatment
is of particular importance for avoiding the occurrence of spurious 
instabilities in neutron-star dynamical simulations. For this reason, unified 
EoSs are well suited for neutron-star modelling. In turn, by comparing these 
models to astrophysical observations, one can extract valuable information on 
the EoS of INM (see, e.g., Ref.~\cite{lat12} for a recent review; see also 
Section III of this volume). 

We describe our recent HFB models in Section~\ref{hfb}, showing how they can be used
to extract the symmetry coefficient $J$ from the mass data. In Section~\ref{nstar}
we apply these models to a unified treatment of neutron stars, calculating in 
particular the composition and the mass-radius relation, with an assessment of the
impact of the symmetry energy. Our conclusions are summarized in 
Section~\ref{concl}. 

\section{The Skyrme-Hartree-Fock-Bogoliubov mass models}
\label{hfb}

Our HFB mass models are intended to facilitate the reliable extrapolation of 
nuclear-mass data out to the highly neutron-rich environments of astrophysical
interest, such as neutron stars and the gravitationally collapsing cores of 
supernovae. As such, they are well adapted to the extraction of information on
symmetry energy from both nuclear data and the observation of neutron stars. 
They likewise enable us to investigate the implications of our (incomplete)
knowledge of symmetry energy for the properties of neutron stars and supernova 
cores. 

In this paper we shall refer only to our latest mass models, HFB-19 to HFB-26,
along with their underlying density functionals, BSk19 to BSk26, 
respectively~\cite{gcp10,gcp13}. All are based on the 16-parameter generalized 
form of Skyrme force given in Eq.~(1) of Ref.~\cite{gcp10}, the unconventional
feature of which is the appearance of so-called $t_4$ and $t_5$ terms, 
density-dependent generalizations of the usual $t_1$ and $t_2$ terms, 
respectively. The full formalism for this generalized Skyrme force is 
presented in the Appendix of Ref.~\cite{cgp09}, but note that we now drop all 
the terms quadratic in the spin-current tensor and their time-odd counterpart
from the Hamiltonian density, for the reasons explained in Ref.~\cite{cg10}. 
The parameters of this form of force were determined
primarily by fitting measured nuclear masses, which were calculated with the 
HFB method. For this it was necessary to supplement the Skyrme forces with a
microscopic pairing force (5 parameters), phenomenological Wigner terms (4
parameters) and correction terms for the spurious collective energy (5 
parameters). 

However, with a view to enhancing the reliability of the extrapolations to all
parts of neutron stars, and to other neutron-rich systems as well, in fitting 
the mass data we simultaneously constrained the Skyrme force to fit the 
zero-temperature EoS of homogeneous neutron matter (NeuM), 
as determined by many-body calculations with realistic two- and three-nucleon
forces. Actually, several realistic calculations of the EoS of NeuM have been
made, and, while they all agree fairly closely at nuclear and subnuclear
densities, at the much higher densities that can be encountered towards the
center of neutron stars they differ greatly in their stiffness, and there are
very few data, either observational or experimental, to discriminate between
the different possibilities. We therefore considered three different 
constraining EoSs of NeuM, as follows. The softest is the one that we label 
FP~\cite{fp81} in Ref.~\cite{gcp10}, the one of intermediate
stiffness is the ``A18 + $\delta\,v$ + UIX$^*$" EoS~\cite{apr98}, which we 
label as APR, while our stiffest constraining EoS is the one labeled ``V18" in 
Ref.~\cite{ls08}, which we refer to as LS2 in Ref.~\cite{gcp10}. 

In constraining our forces to the EoS of NeuM we paid particular attention
to the high densities appropriate to neutron-star cores. At the same time
all of the models of this paper were fitted to one value or another of the
symmetry coefficient $J$ in the range 29 to 32 MeV. That we were able to 
adjust the symmetry energy at supernuclear densities more or less 
independently of
the symmetry energy at nuclear densities is a consequence of the high 
flexibility of the 16-parameter Skyrme force that we have taken.   

\subsection{The 2010 Models}  

The three density functionals of Ref.~\cite{gcp10}, BSk19, BSk20 and BSk21,
were constrained by the FP, APR and LS2 EoSs of NeuM, respectively, i.e., in 
ascending order of high-density stiffness. All had the value of 30 MeV for
the symmetry coefficient $J$ imposed on them, and all of them were fitted to
the 2149 measured masses of nuclei with $N$ and $Z \ge$ 8 given in the 2003 
AME~\cite{ame03}, that being the latest available at the time. The rms
deviation of these mass fits was 0.58 MeV in all three cases. We showed in 
Fig.~1 of Ref.~\cite{gcp10} how well each of the three models reproduces its 
realistic ``target'' EoS of NeuM. 

\begin{table}
\caption{Maximum neutron-star mass $\mathcal{M}_{max}$ (in units of the mass of the 
Sun $\mathcal{M}_{\odot}$) for different models.}
\label{tab1}
\begin{tabular}{|c|c|}
\hline\noalign{\smallskip}
Force&  $\mathcal{M}_{max}/\mathcal{M}_{\odot}$\\
\noalign{\smallskip}\hline\noalign{\smallskip}
BSk19&  1.86  \\
BSk20&  2.15 \\
BSk21&  2.28 \\
BSk22&  2.26 \\
BSk23&  2.27 \\
BSk24&  2.28  \\
BSk25&  2.22  \\
BSk26&  2.14   \\
\noalign{\smallskip}\hline
\end{tabular}
\end{table}

One might expect that the stiffer the constraining EoS the greater the maximum
neutron-star mass that can be supported, and this was indeed found to be the
case~\cite{cha11} on solving the Tolman-Oppenheimer-Volkoff (TOV) 
equations~\cite{tol39,ov39} and imposing causality, as can be seen in the first 
three lines of Table~\ref{tab1}. 
As a matter of fact, the maximum neutron-star masses thus obtained are very 
close to those found with the corresponding constraining microscopic equation 
of state of NeuM. 
After the publication of our 2010 paper~\cite{gcp10} the
neutron stars PSR J1614$-$2230 and PSR J0348+0432 were shown to have masses of
1.97 $\pm$ 0.04 $\mathcal{M}_{\odot}$~\cite{dem10} and 2.01 $\pm$ 0.04 
$\mathcal{M}_{\odot}$~\cite{ant13}, respectively. 
This means that the neutron-star matter EoS obtained with BSk19 is 
definitely too soft, and it must be discarded.
This does not necessarily imply that the NeuM EoS has to be much stiffer 
than that of FP, because the core of neutron stars may contain non-nucleonic 
particles~(see, e.g., Ref.~\cite{cfpg13}); we do not consider this possibility
any further in this paper. 

\subsection{The 2013 Models}

In refitting to the 2353 measured masses of nuclei having $N$ and $Z \ge$ 8 in
the 2012 AME~\cite{ame12} we took the opportunity of imposing different values
of the symmetry coefficient $J$ on our fits. We generated in all five new
parameter sets, BSk22 to BSk26, along with the corresponding mass tables,
labeled HFB-22 to HFB-26, respectively~\cite{gcp13}. BSk22 to BSk25 were fitted
to $J$ = 32, 31, 30 and 29 MeV, respectively, and were all constrained to LS2, 
while BSk26 was also fitted to $J$ = 30 MeV but under the APR constraint.

The rms and mean (data - theory) values of the deviations between the measured
masses and the predictions for the five new models are given in the first and
second lines, respectively, of Table~\ref{tab2}. The next two lines of this
table show the corresponding deviations for the subset consisting of the most
neutron-rich measured nuclei, here taken as those with a neutron separation
energy $S_n \le $ 5.0 MeV (there are 257 nuclei in this subset). All five
models display, not surprisingly, some deterioration as we move into the 
neutron-rich region. From the first line we see that the parameter sets BSk24 
($J$ = 30 MeV) and BSk25 ($J$ = 29 MeV) give the best global fits of all the 
new models, and in fact are better than any of our previous models. However, 
line 3 of Table~\ref{tab2} shows that the deterioration of BSk25 on moving into
the neutron-rich region is much stronger than for BSk24: for all the other 
models the performance in the neutron-rich region correlates fairly well with 
the global performance. Thus the apparent high performance of this model in the
global fit should be interpreted with caution; other defects of model BSk25 are
discussed in Ref.~\cite{gcp13}. Looking at BSk22 and BSk23, we see from both 
lines 1 and 3 that $J$ = 31 MeV works less well than either 29 or 30 MeV, while
$J$ = 32 MeV is still more strongly disfavoured.

Comparing BSk24 and BSk26 shows that for $J$ = 30 MeV the high-density LS2 
constraint (BSk24) gives slightly better fits than APR (BSk26). It must be 
stressed, however, that this discrimination in favour of LS2 as the constraining
EoS of NeuM relates only to nuclear and subnuclear densities. In particular, we
should not conclude that nuclear masses are telling us something about the EoS 
of NeuM at the higher densities found in neutron-star cores, since it is 
conceivable that our 16-parameter Skyrme form could be generalized still further
in such a way that the EoS at high densities
was changed without affecting the fit to nuclear masses.

Overall, the clearest conclusion that can be drawn from Table~\ref{tab2} is
that model BSk22 is the worst performing of all our models, ruling out
$J$ = 32 MeV. There are also very strong indications that
$J$ = 29 or 30 MeV (the latter in both its LS2 and APR forms) are to be
preferred to $J$ = 31 MeV, although we have expressed some concerns
with regards to $J$ = 29 MeV, i.e., to BSk25. In any case, it will be seen
that the HFB models favour a value of $J$ considerably higher than what we 
found from the simple drop model of Eq.~(\ref{1}),
illustrating thereby the importance of using a model that takes account of
the fine details on nuclear structure.

\begin{table*}
\caption{Rms ($\sigma$) and mean ($\bar{\epsilon}$) deviations between data
and predictions for the 2013 models. The first pair of lines refers to all
the 2353 measured masses $M$ that were fitted \cite{ame12}, and the second pair
to the masses $M_{nr}$ of the subset of 257 neutron-rich nuclei 
($S_n \le $ 5.0 MeV).}
\label{tab2}
\begin{tabular}{|c|ccccc|}
\hline\noalign{\smallskip}
&HFB-22&HFB-23&HFB-24&HFB-25&HFB-26\\
\noalign{\smallskip}\hline\noalign{\smallskip}
$\sigma(M)$ {\scriptsize [MeV]}               &0.629 &0.569 &0.549 &0.544  & 0.564  \\$\bar{\epsilon}(M)$ {\scriptsize [MeV]}       &-0.043&-0.022&-0.012&0.008  & 0.006  \\
$\sigma(M_{nr})$ {\scriptsize [MeV]}          &0.817 &0.721 &0.702 &0.791  & 0.749  \\
$\bar{\epsilon}(M_{nr})$ {\scriptsize [MeV]}  &0.221 & 0.090&0.011 &0.023  & 0.230  \\
\noalign{\smallskip}\hline
\end{tabular}
\end{table*} 

\begin{figure}
\includegraphics[scale=0.3]{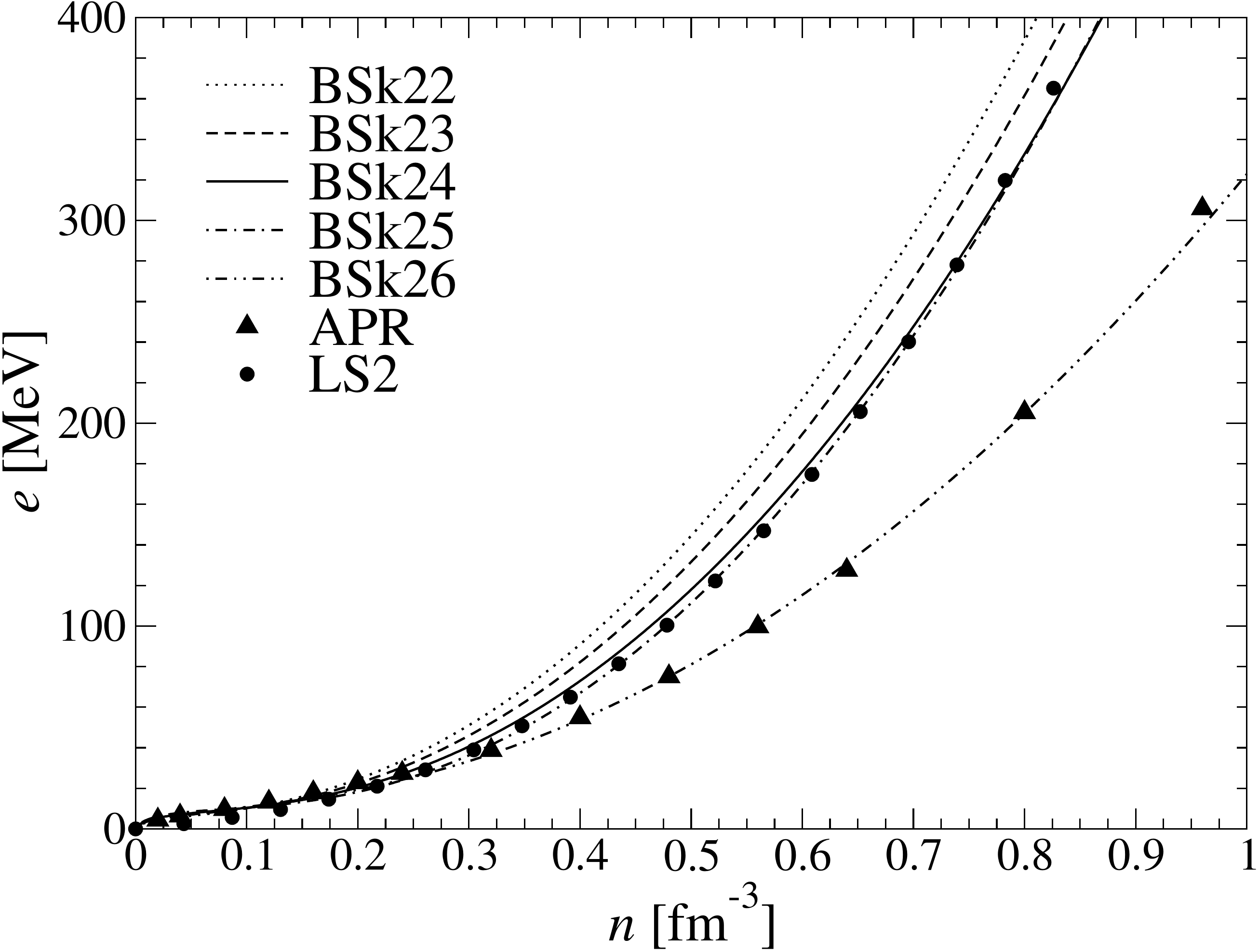}
\caption{Zero-temperature EoSs for neutron matter with models BSk22-26.
APR is the realistic EoS ``A18 + $\delta\,v$ + UIX$^*$" in Ref.~\cite{apr98}.
LS2 is the realistic EoS ``V18" in  Ref.~\cite{ls08}.}
\label{neum}
\end{figure}

\begin{figure}
\includegraphics[scale=0.3]{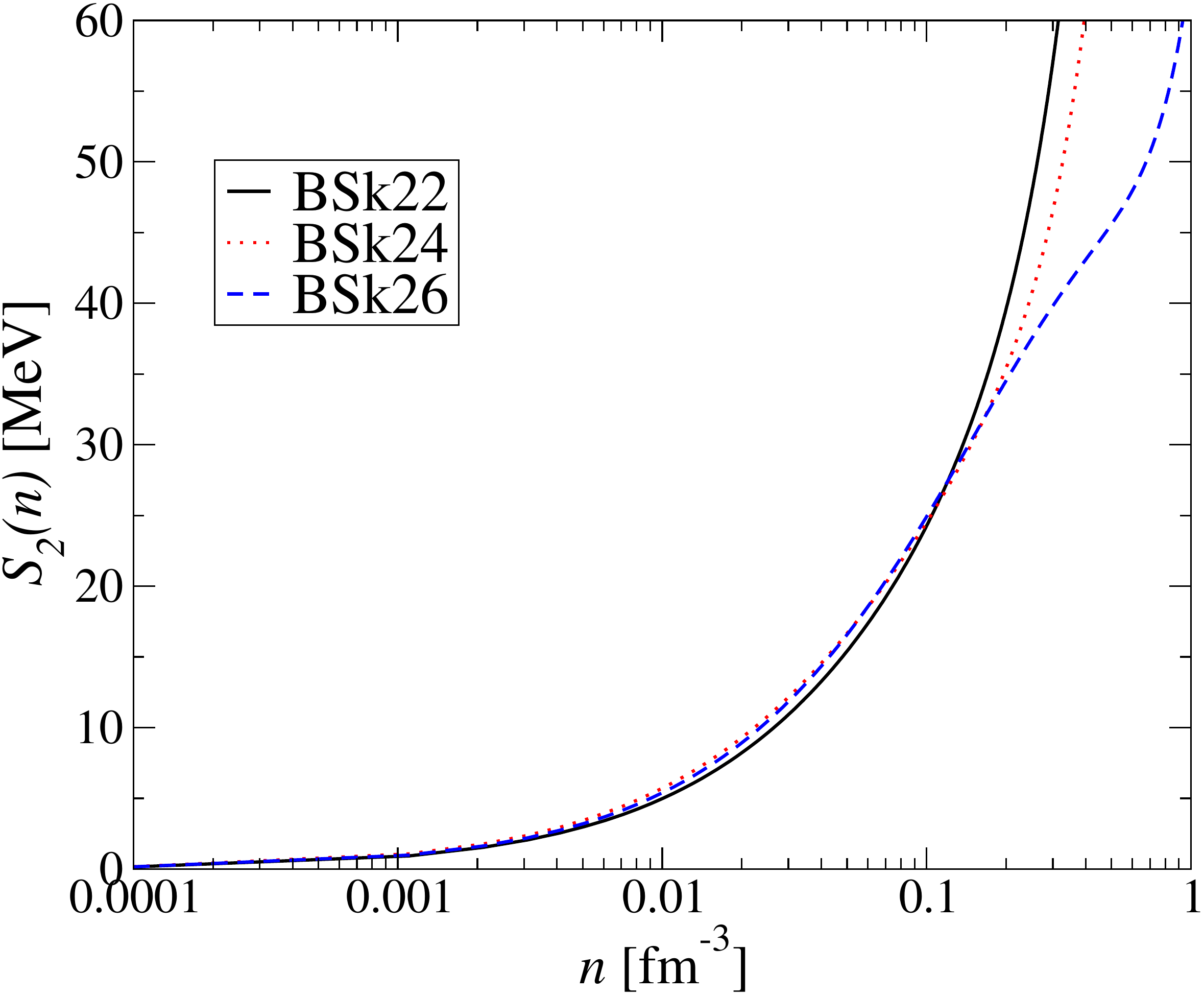}
\caption{Symmetry energy $S_2(n)$ for models BSk22, BSk24 and BSk26.}
\label{esym}
\end{figure}

In Fig.~\ref{neum} we show the neutron-matter EoSs of our five models; it 
will be seen how well they fit their ``target" EoSs. Calculating symmetric INM
for the five models enables us to calculate the symmetry energy $S_2(n)$, as
given by Eq.~(\ref{6}). We show in Fig.~\ref{esym} the variation of $S_2(n)$ with
density for models BSk22 ($J$ = 32 MeV, constrained by LS2), BSk24 ($J$ = 30 MeV, 
constrained by LS2) and BSk26 ($J$ = 30 MeV, constrained by APR); with this
choice of models we are able to sample the influence of both the symmetry
energy at nuclear densities and the high-density behavior of NeuM. We do not
show the corresponding curves for symmetric INM, but they are all remarkably 
similar (see the discussion of this point in Section IIIA of Ref.~\cite{gcp10}).

{\it Neutron skins.} The most extensive set of measurements of neutron-skin
thicknesses using a given method is that of Ref.~\cite{jast04}, which used
antiproton scattering to measure 26 nuclei. In Table~\ref{tab3} we show the rms
deviations $\sigma_{rms}$ between our models and experiment, the mean
deviations $\bar{\epsilon}$ and the model error $\sigma_{mod}$ of M\"{o}ller
and Nix~\cite{mol88}. This last quantity provides a more reliable method of
assessing the relative performance of different models, especially when the
experimental errors are large, as in the present case. In the first three 
columns of this table we show the results for the full set of 26 nuclei, while 
in the next three columns we consider only the 10 nuclei for which the 
experimental errors are 0.04 fm or less. We see that all three deviations, for 
both the complete set of data and the subset, show that models BSk24 and BSk26 
are in better agreement with experiment than are the other models, thereby
supporting the conclusion drawn from masses favoring $J$ = 30 MeV.  

\begin{table*}
\caption{Rms, mean and model deviations between 2013 models and the
neutron-skin thickness measurements of Ref.~\cite{jast04}. The first three columns relate to the
complete set of 26 nuclei, the last three colums to the set of 10 nuclei with
the lowest experimental error.}
\label{tab3}
\vspace{.5cm}
\begin{tabular}{|c|ccc|ccc|}
\hline\noalign{\smallskip}
&$\sigma_{rms}$ (26)&$\bar{\epsilon}$ (26)&$\sigma_{mod}$ (26) &$\sigma_{rms}$
(10)&$\bar{\epsilon}$ (10)&$\sigma_{mod} (10)$\\
\noalign{\smallskip}\hline\noalign{\smallskip}
BSk22&0.0495&-0.0266&0.0205&0.0429&-0.030&0.0244\\
BSk23&0.0447&-0.0142&0.0090&0.0342&-0.017&0.0128\\
BSk24&0.0437&-0.0031&0.0047&0.0270&-0.0030&0.0087\\
BSk25&0.0469&0.0088&0.0170&0.0277&0.011&0.0194\\
BSk26&0.0415&-0.0038&0.0044&0.0265&-0.0040&0.0084\\
\noalign{\smallskip}\hline
\end{tabular}
\end{table*}

\begin{table}
\centering
\caption{$J$ and higher-order symmetry coefficients for the 2013 models.}
\label{tab4}
\vspace{.5cm}
\begin{tabular}{|c|ccccc|}
\hline\noalign{\smallskip}
                                 &BSk22  &BSk23  &BSk24  &BSk25   &BSk26    \\
\noalign{\smallskip}\hline\noalign{\smallskip}
$J$ {\scriptsize [MeV]}          &32.0   &31.0   &30.0   &29.0    &30.0 \\ 
$L$ \scriptsize [MeV]            &68.5   &57.8   &46.4   &36.9    &37.5 \\
$K_{sym}$\scriptsize [MeV]       &13.0   &-11.3  &-37.6  &-28.5   &-135.6  \\
$K^{\prime}$ \scriptsize [MeV]   &275.5  &275.0  &274.5  &316.5   &282.9   \\
\noalign{\smallskip}\hline
\end{tabular}
\end{table}

{\it Correlations between $J$ and higher-order symmetry coefficients.} In
Table~\ref{tab4} we show for each of our 2013 models the values of $J$ and the
higher-order symmetry coefficients, $L$ and $K_{sym}$ defined in 
Eq.~(\ref{5b}). Although we have concluded that $J$ must have a value close to 
29 or 30 MeV, with 32 MeV definitely excluded, it is still of interest to see 
what happens to $L$ and $K_{sym}$ when we vary $J$ over the full range of 29 to
32 MeV. Table~\ref{tab4} shows that in our models $L$ is strongly correlated 
with $J$, confirming what was first established at least 35 years 
ago~\cite{fpr78}. In fact, when plotted in the $J - L$ plane all our models 
fall within the elliptical region labeled ``Nuclear masses" in Fig. 12 of
Ref.~\cite{lat12}. Actually, if we compare the $L$-values for models BSk24 and
BSk26, both of which have $J$ = 30 MeV, we see that there is some dependence
on the constraining EoS, the softer EoS leading, not surprisingly, to a lower
value of $L$. 

The influence of the constraining EoS is seen to be much stronger in the case 
of $K_{sym}$, the correlation with $J$ (or $L$) being quite weak;
clearly, a determination of $L$ will not suffice to determine $K_{sym}$, 
contrary to the assertion of Ref.~\cite{dong12}. This apparent correlation 
between $K_{sym}$ and the constraining EoS of NeuM
might tempt one to conclude that a measurement of $K_{sym}$  will suffice 
to determine the high-density EoS of NeuM. Once again, however, we must
realize that our 16-parameter Skyrme form could be generalized still further, 
making it possible to change the EoS at high densities without affecting $K_{sym}$.

\begin{figure}
\includegraphics[scale=0.3]{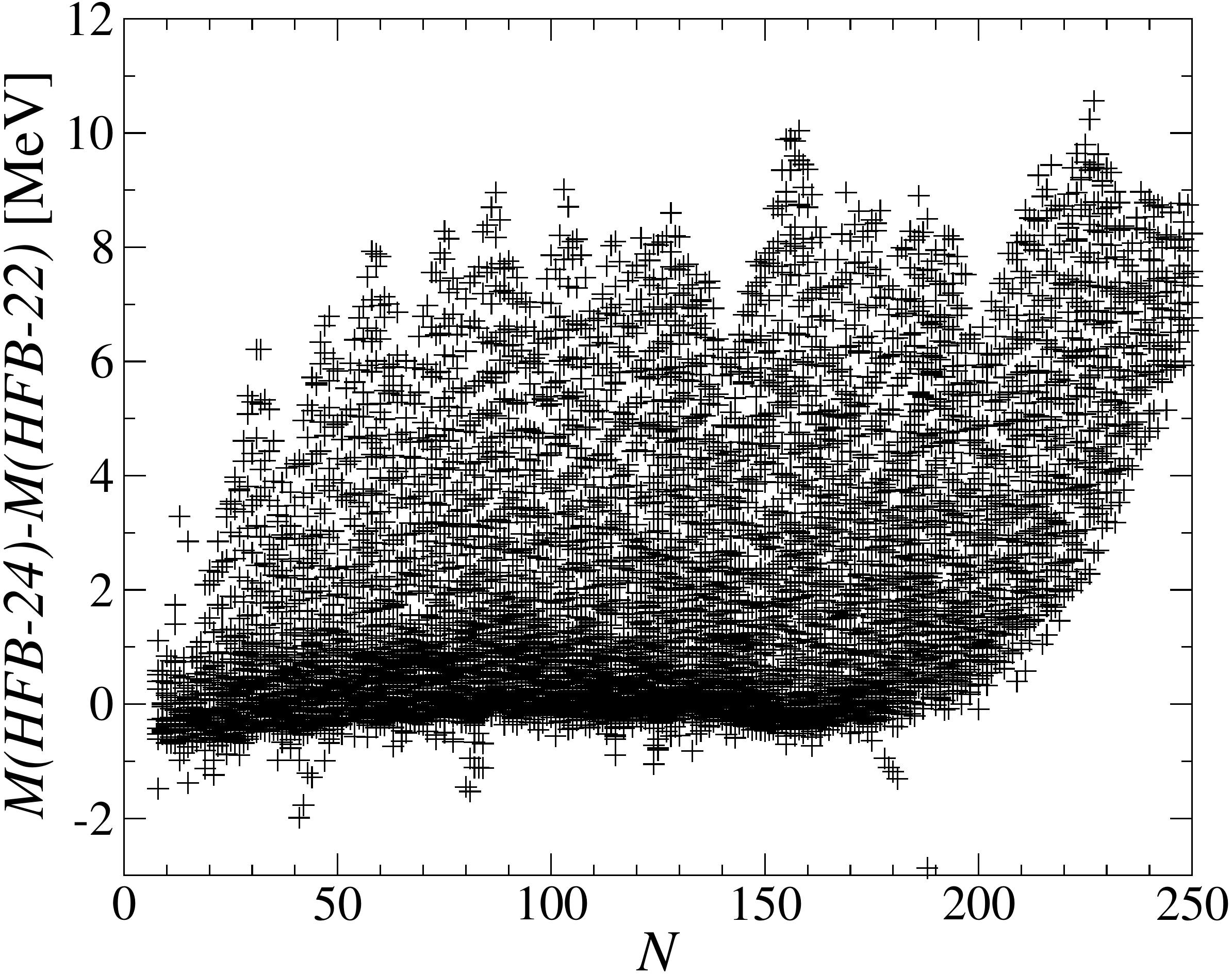}
\caption{Differences between HFB-22 and HFB-24 mass predictions for all nuclei
between the drip lines.}
\label{extrap}
\end{figure}

{\it Extrapolation of masses far from the stability line.} We show in 
Fig.~\ref{extrap}, for all the nearly 8500 nuclei with $8 \le Z \le 110$ lying
between the proton and neutron drip lines, the deviations between mass models
HFB-22 ($J$ = 32 MeV) and HFB-24 ($J$ = 30 MeV). Discrepancies of up to 10 MeV 
will be seen, and it is noteworthy that the lower the $J$-value the larger the 
masses predicted far from the stability line. This behaviour is somewhat
counter-intuitive, but it is by no means peculiar to the present forces, having been
remarked before~\cite{sg05}. The explanation can be found in Fig.~\ref{esym}, where
it will be seen that the average value of the symmetry energy over the nuclear 
volume is smaller the greater $J$ (this argument remains valid whether one takes
$S_1$ or $S_2$). In the liquid-drop model this average
symmetry energy corresponds to the coefficient $(J + a_{ss}A^{-1/3})$ of the $I^2$ 
term in Eq.~(\ref{1}), so that we should expect a strong anticorrelation between $J$
and the surface-symmetry coefficient $a_{ss}$. This we have verified to be the case
by fitting the mass data for different fixed values of $J$, and we have found in
fact that $(J + a_{ss}A^{-1/3})$ decreases with increasing values of J for all
realistic values of $A$.   

\begin{figure}
\includegraphics[scale=0.3]{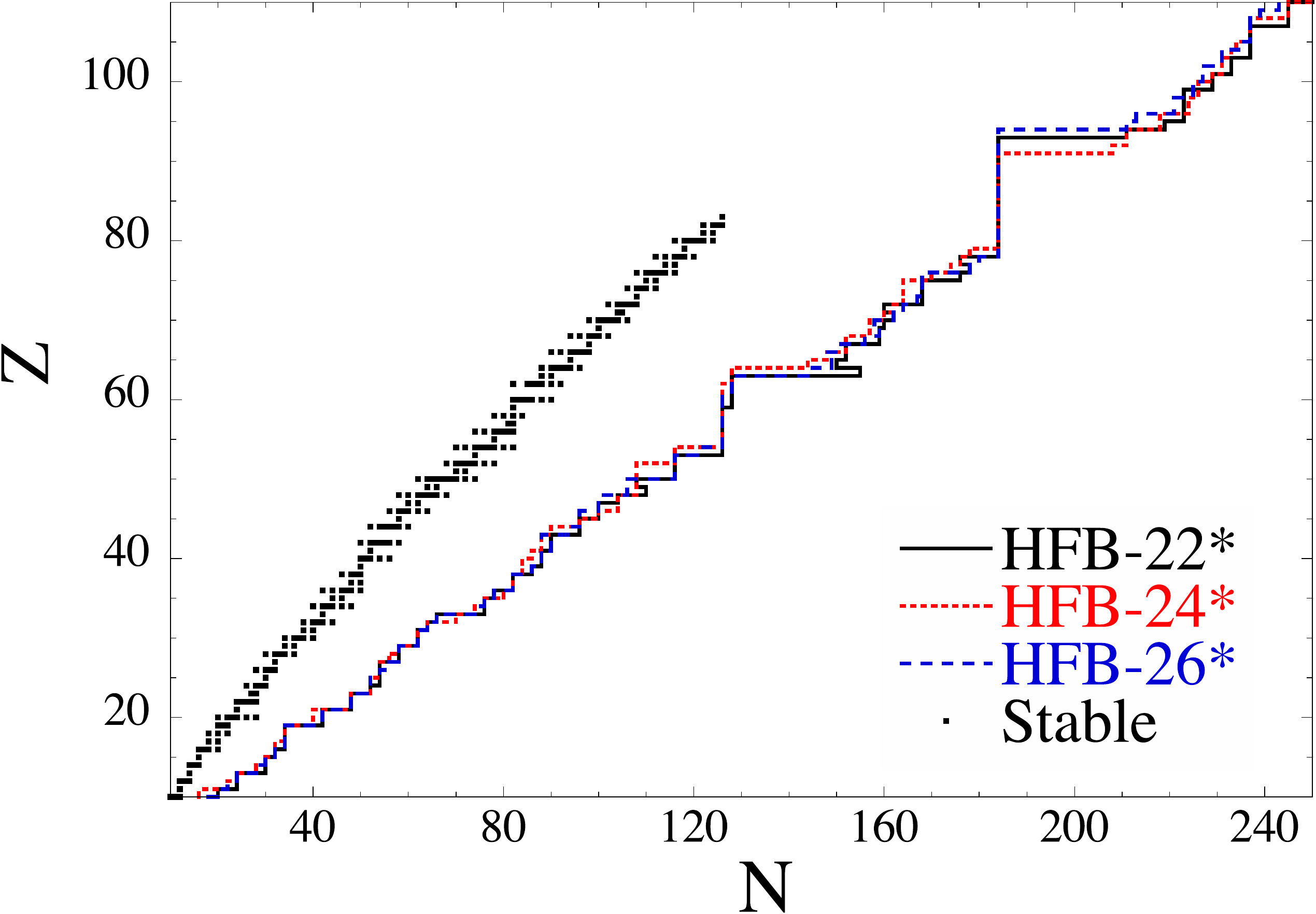}
\caption{Neutron drip lines for models HFB-22, HFB-24 and HFB-26.}
\label{drip}
\end{figure}

{\it Neutron drip lines.} We show in Fig.~\ref{drip} the neutron drip 
lines for mass models HFB-22, HFB-24 and HFB-26. 
Recalling that the neutron drip line represents for each value of $Z$ the lowest
value of $N$ for which the neutron separation energy $S_n$ vanishes, it will
be understood that the position of the drip line for any given model will
be highly sensitive to the small random numerical errors that are inevitable
in computing. Thus in calculating the drip lines we first apply the noise 
filter described in the Appendix, denoting the model thereby generated by
HFB-22$^*$, etc. (nowhere else in this paper, or in any any other that we have
published so far, do we use these noise-filtered models).  

Given the sensitivity to $J$ of the masses of nuclei far from the stability line 
seen in Fig.~\ref{extrap}, it is somewhat surprising that the different drip lines
shown in Fig.~\ref{drip} lie so close to each other. The reason is that the
$S_n$ of these nuclei are much less sensitive to $J$ than are the masses.   

\section{Implications of HFB models for neutron stars}
\label{nstar}

We have calculated the composition and EoS in all three regions of
neutron stars (outer crust, inner crust and core) for the three typical 2013 
models: BSk22 ($J$ = 32 MeV, constrained by LS2), BSk24 ($J$ = 30 MeV, 
constrained by LS2) and BSk26 ($J$ = 30 MeV, constrained by APR). The outer 
crust is calculated as in Ref.~\cite{pgc11}, and the inner crust as in
Ref.~\cite{pcgd12}, except that we now take into account proton pairing 
correlations in the BCS approximation along the lines of Ref.~\cite{pea91}. 
The calculation of the neutron-star matter of the core is, in view of the homogeneity, 
essentially exact: analytic expressions for both the nuclear and leptonic (electrons and
muons) contributions are given in Ref.~\cite{cgp09}, and minimization with
respect to the proton fraction $Y_p$ and the muonic fraction of the leptons can
be carried through to any required degree of accuracy.

\begin{figure}
\includegraphics[scale=0.3]{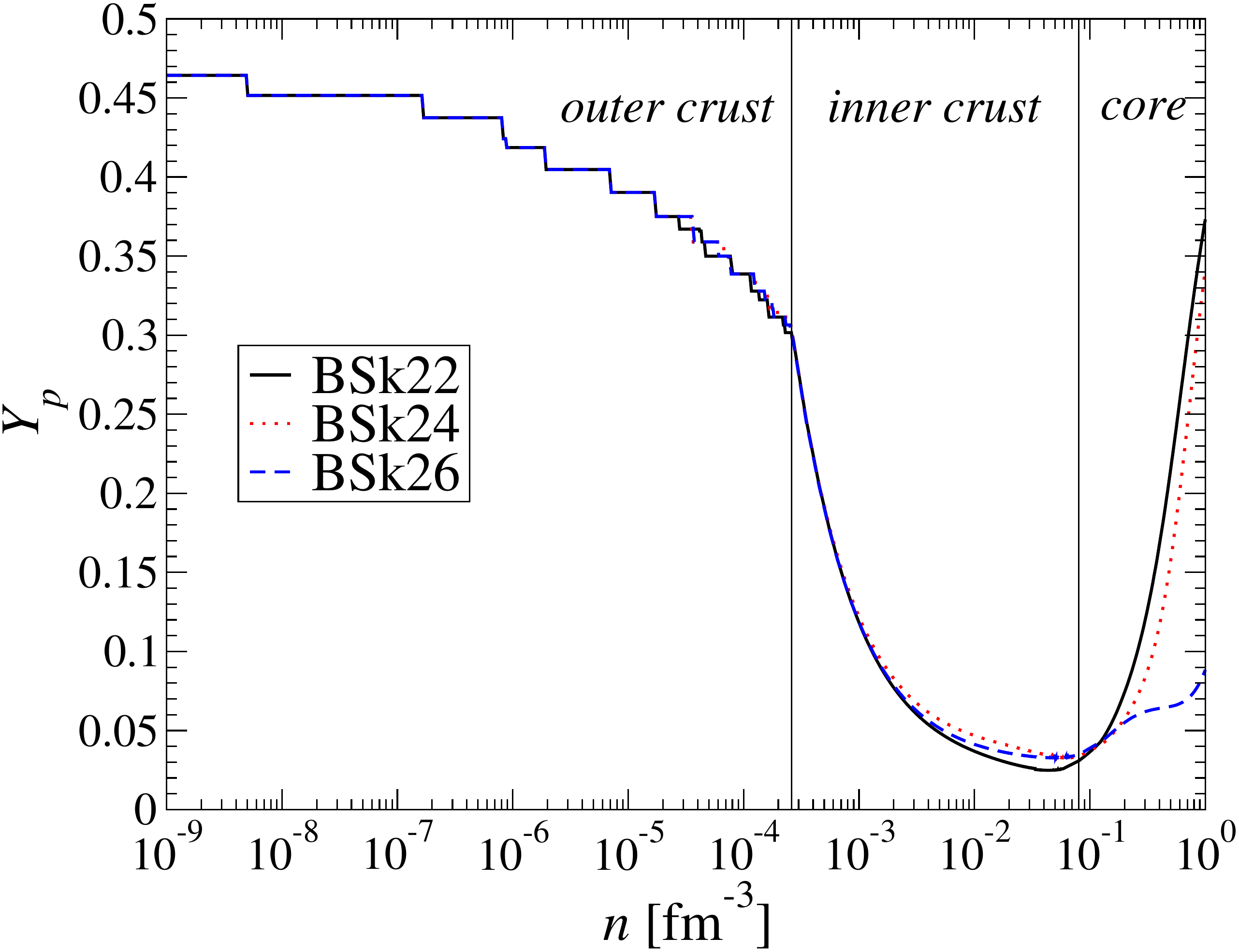}
\caption{Proton fraction $Y_p$ as a function of mean local density in neutron stars
for models BSk22, BSk24 and BSk26.}
\label{yp}
\end{figure}

The variation of $Y_p$ with density over the entire star that we found for
the three models is shown in Fig.~\ref{yp}, where the boundaries between the three
regions are indicated. As we move below the surface the energy of the degenerate
electrons dominates and the protons begin to capture electrons, i.e., $Y_p$ 
decreases. However, at some point in the core the process is reversed and $Y_p$ 
begins to increase with increasing density. That is, the neutronisation of protons
gives way to the protonisation of neutrons through beta decay,  
because the nuclear symmetry energy rises so rapidly 
that it dominates the degenerate electrons. All three models behave very
similarly as far as $Y_p$ is concerned, but differences can be discerned in the
higher density range of Fig.~\ref{yp}, and in fact they are seen to be 
correlated with the differences in the symmetry energy shown in Fig.~\ref{esym}. 

The proton fraction $Y_p$ is not the only composition variable of interest. 
For many purposes it is important to know the values of the proton number $Z$ 
for the nuclei of the outer crust; we do not show them here, but we will publish 
them elsewhere. The situation concerning the $Z$-values of the clusters in the 
Wigner-Seitz cells of the inner crust is very simple: we have $Z$ = 40 throughout
the inner crust for all models, except for BSk22, where $Z = 20$ is favoured for 
densities higher than 0.035 nucleons per cubic fermis. (Actually, the existence of 
strong shell effects for $Z$ = 20 and 40 is evident throughout the inner 
crust; the absence of $Z$ = 28 might be taken as an indication of a suppression
of the spin-orbit coupling.) 

{\it Maximum mass.} Using the EoSs that we have calculated, 
we solve the TOV equations to obtain the maximum mass 
of non-rotating stars for each of the five new
density functionals of 2013, and can thereby complete Table~\ref{tab1}. 
The numerical calculations have been done using the LORENE library
(\texttt{http://www.lorene.obspm.fr}, see also Ref.~\cite{gourgoulhon2010}). 
We see that all our new models are compatible with the existence of the massive 
neutron stars PSR J1614$-$2230 and PSR J0348+0432. It turns out that the 
maximum mass is essentially the same for all models constrained by the same 
EoS of NeuM, and, in particular, seems to be independent of the symmetry 
energy.

\begin{figure}
\includegraphics[scale=0.3]{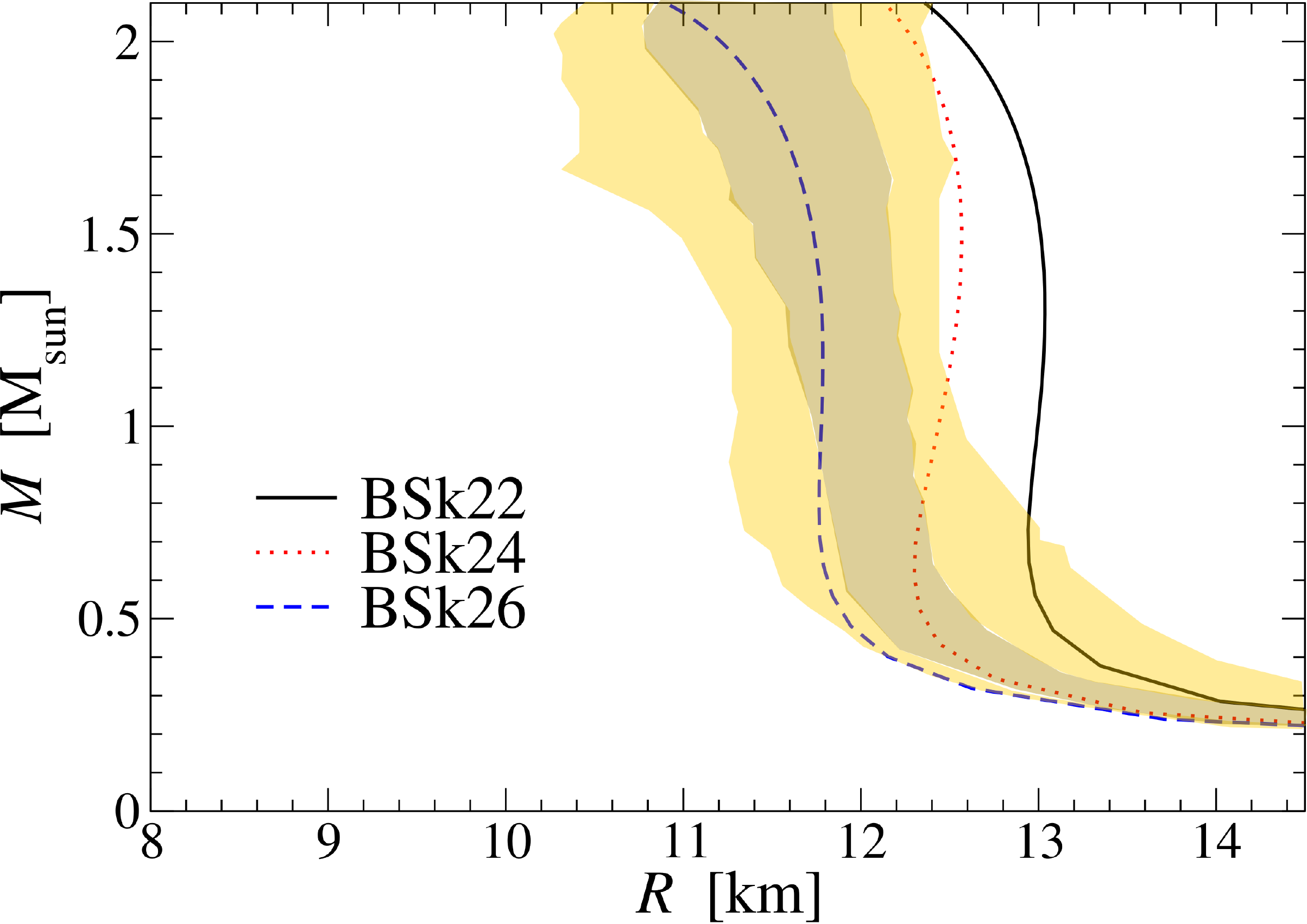}
\caption{Mass-radius relation of neutron stars for models BSk22, BSk24 and BSk26.
The dark (light) shaded regions correspond to the 1-$\sigma$ (2-$\sigma$) probability
distributions implied by observation, as shown in the upper right panel of Fig.~9
of Ref.~\cite{stei10}.}

\label{mr}
\end{figure}

{\it Mass-radius relation.} In a similar way we calculated the mass-radius
relation for non-rotating stars with each of the models BSk22, BSk24 and BSk26. 
Our results are compared with the constraints obtained from astrophysical 
observations of three X-ray bursters of type I 
with photospheric radius expansion, and 
three transient low-mass X-ray binaries in globular clusters~\cite{stei10}.  
It will be seen in Fig.~\ref{mr} that model BSk22 fails completely this
test, BSk24 is at the limit of acceptability, while BSk26 is highly 
satisfactory. Thus, once again, we have support for preferring
$J$ = 30 MeV over 32 MeV, while the APR EoS of NeuM seems to be somewhat
favoured over the stiffer LS2 EoS.

{\it Direct Urca.} The density above which the direct Urca process is allowed is 
indicated in Table~\ref{tab5}, as well as the corresponding neutron-star mass for 
models BSk22, BSk24 and BSk26. All models predict the onset of the direct Urca process
at high enough densities. But for the model BSk26, this process occurs at densities 
exceeding the central density in the most massive stable neutron star. On the other hand, 
the pulsar in CTA1, the transiently accreting millisecond pulsar SAX J1808.4$-$3658 and 
the soft X-ray transient 1H 1905$+$000 appear to be too cold, thus suggesting that these 
neutron stars cool very fast via the direct Urca process~\cite{jonker2007,heinke2009,page09}. 
Moreover, the low luminosity from several young supernova remnants likely to contain a still 
unobserved neutron star~\cite{kapl04,kapl06} provide further evidence for a direct Urca 
process~\cite{page09,sy08}. If this interpretation is correct, model BSk26 would be 
ruled out. In addition, the model BSk22 is found to be incompatible with the constraint of 
Kl{\"a}hn et al.~\cite{klahn2006} that no direct Urca process should occur in neutron stars 
with typical masses in the range $\mathcal{M} \sim 1 - 1.5$~$\mathcal{M}_\odot$, 
leaving thereby BSk24 as the only model consistent with both constraints.

\begin{table}
\centering
\caption{Threshold density for the direct Urca process to occur, and corresponding 
neutron-star mass for models BSk22, BSk24 and BSk26.}
\label{tab5}
\vspace{.5cm}
\begin{tabular}{|c|cc|}
\hline\noalign{\smallskip}
        & $n_{du}$  &  $\mathcal{M}_{du}/\mathcal{M}_{\odot}$   \\
\noalign{\smallskip}\hline\noalign{\smallskip}
BSk22   & 0.33   & 1.14 \\ 
BSk24   & 0.45   & 1.59 \\
BSk26   & 1.46   & -  \\
\noalign{\smallskip}\hline
\end{tabular}
\end{table}

We do not show here the EoSs that we have calculated and that are implicit in 
all the results of this section, but will publish them elsewhere.

\section{Conclusions}
\label{concl}

We have described the main features of our most recent HFB mass models, which
have been fitted to the mass data of the 2012 AME. Our preferred model, HFB-24,
fits the 2353 measured masses of nuclei with
$N$ and $Z \ge$ 8 with an rms deviation under 0.55 MeV. A symmetry coefficient of
$J$ = 30 MeV is favoured, with strong indications against a value of 32 MeV.

These mass models, taken with their underlying functionals, permit a unified
treatment of all regions of neutron stars. We have calculated with each of our 
models the composition and EoS of neutron stars. Solving then the TOV equations we
were able to calculate the mass-radius relation for each model, which, when compared
with the observational constraints, strongly favours $J$ = 30 MeV over 32 MeV. 
Moreover, all our models lead to maximum masses greater than the heaviest observed 
neutron stars.

\begin{acknowledgement}
This work was financially supported by the F.R.S.-FNRS (Belgium) and the NSERC (Canada).
\end{acknowledgement}

\appendix
\renewcommand{\theequation}{A\arabic{equation}}
\setcounter{equation}{0}

\section{Noise-filtering of the calculated model masses}
\label{smooth}

\begin{figure}
\includegraphics[scale=0.3]{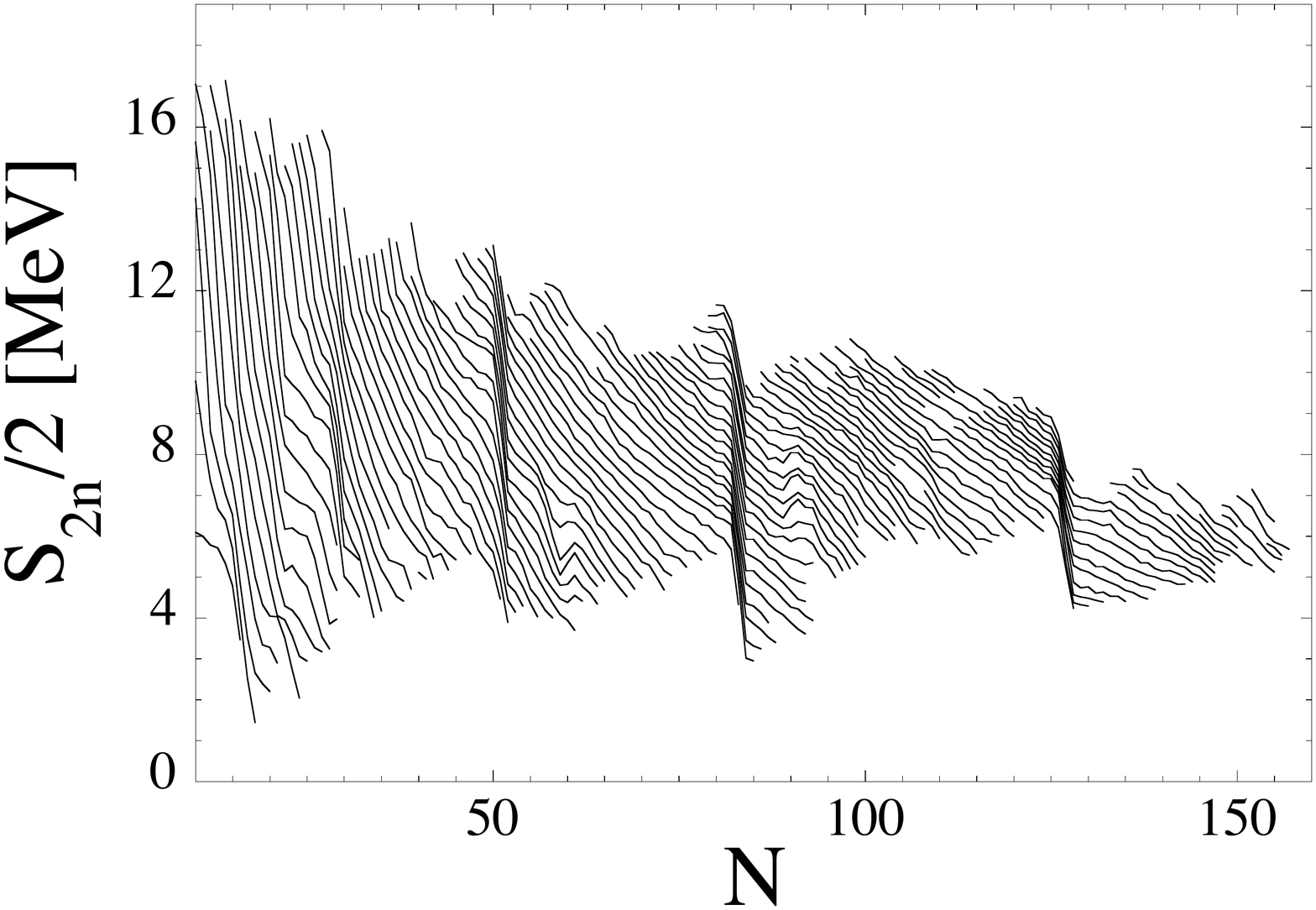}
\caption{Two-neutron separation energies for all isotopic chains with known masses}
\label{fig_A1}
\end{figure}

\begin{figure}
\includegraphics[scale=0.3]{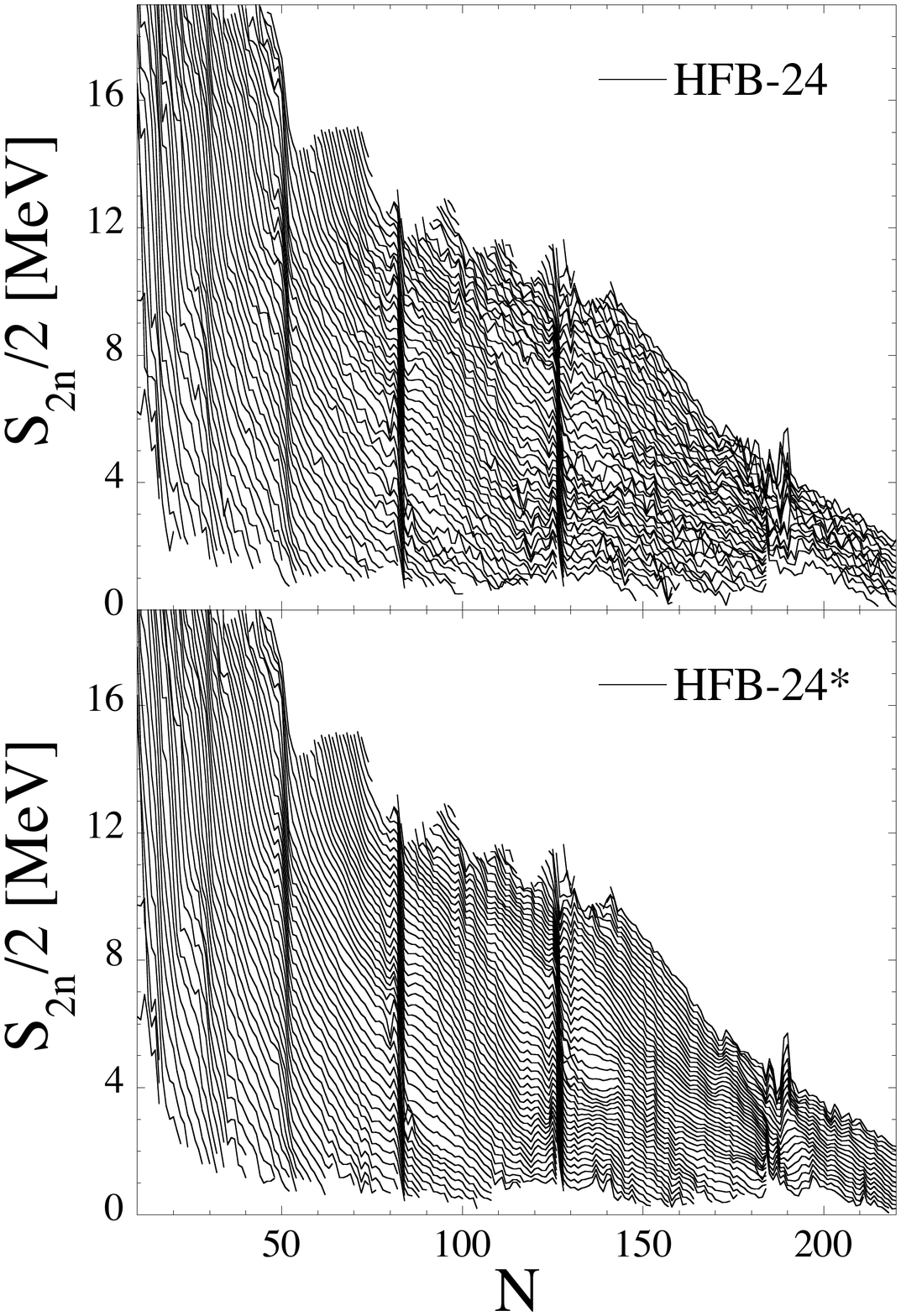}
\caption{$S_{2n}/2$ surfaces for HFB24 mass table before (upper panel) and
after (lower panel) filtering the masses by the ``21GK" method described in the
text. }
\label{fig_A2}
\end{figure}

In Fig.~\ref{fig_A1} we show the experimental values \cite{ame12} of the two-neutron separation 
energy $S_{2n}$ and in Fig.~\ref{fig_A2} the calculated 
(HFB-24) values. It will be seen in this latter figure 
that despite the excellent global fit to the mass data there are many small
irregularities in the calculated masses, with the different isotopic curves
sometimes touching and even crossing each other, while the experimental curves (Fig.~\ref{fig_A1})
are much smoother and more regularly spaced. This spurious structure is a 
noise that inevitably occurs in the extensive numerical procedures involved in 
solving the HFB equations. 

\begin{figure}
\includegraphics[scale=0.3]{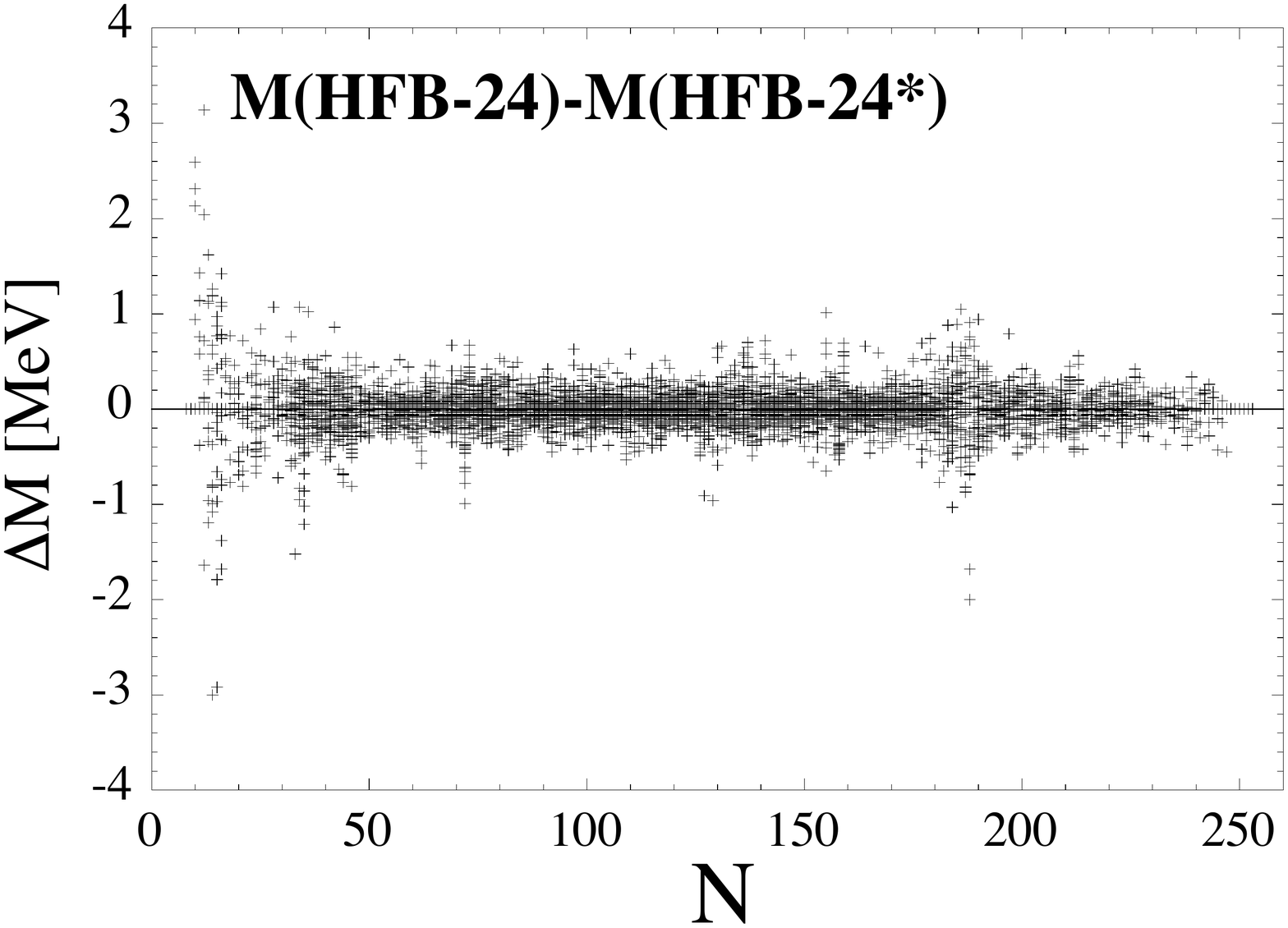}
\caption{Mass differences between the noise-filtered version of the HFB-24 
model and the original version, as a function of neutron number $N$ for the 8509 nuclei 
included in the mass table.}
\label{fig_A3}
\end{figure}

There is, of course, no unambiguous way to filter out this noise, but here we 
describe a procedure that we have developed on the basis of the 21-nucleus 
variation of the Garvey-Kelson relations~\cite{ga69}, abbreviated here as 21GK,
shown by Barea ~{\it et al.}~\cite{ba08} to be well satisfied by the 
mass data of the 2003 AME~\cite{ame03}, the rms deviation for all nuclei with 
$A\ge 16$ being 0.087 MeV. The degree of smoothness displayed by the 21GK
masses is thus much closer to what is seen in the data than in the HFB results.

Our method proceeds reiteratively, with the mass of the nucleus $(Z,N)$ after 
the $i$'th iteration being expressed as 
\begin{align}
\label{eq_gk}
&M_i(Z,N)= \nonumber\\  
&\frac{1}{12}~ [ M_{i-1}(Z+2,N-2)+M_{i-1}(Z+2,N+2) \nonumber \\ 
& +M_{i}(Z-2,N-2) +M_{i}(Z-2,N+2) \nonumber\\ 
&  -2~M_{i-1}(Z+2,N-1)-2~M_{i-1}(Z+2,N+1) \nonumber\\  
&   -2~M_{i-1}(Z+1,N-2)-2~M_{i-1}(Z+1,N+2) \nonumber\\ 
&  -2~M_{i}(Z-1,N-2)-2~M_{i}(Z-1,N+2) \nonumber\\ 
&  -2~M_{i}(Z-2,N-1)-2~M_{i}(Z-2,N+1) \nonumber\\ 
&  +2~M_{i-1}(Z,N+2)+2~M_{i-1}(Z+2,N)\nonumber \\ 
&  +2~M_{i}(Z,N-2)+ 2~M_{i}(Z-2,N) \nonumber\\
&  +4~M_{i-1}(Z,N+1)+4~M_{i-1}(Z+1,N) \nonumber\\ 
& +4~M_{i}(Z,N-1)+4~M_{i}(Z-1,N) ]   \,  ,
\end{align}
where we complete one isotope chain at a time, beginning with the lowest value 
of $Z$, before passing to the next value of $Z$; for each chain we begin with
the lowest value of $N$. Thus to estimate the mass 
$M(Z_0,N_0)$, the masses entering the right-hand side of Eq.~(\ref{eq_gk}) 
correspond to the new iteration for $Z < Z_0$ and also for $Z = Z_0$ provided
$N < N_0$; otherwise we take the previous iteration. The initial iteration, 
$i=0$, corresponds to the original HFB value. We emphasize that this procedure 
is not applied to nuclei if both the neutron and proton number differ by less than
two units from a magic number, thereby keeping the sharp shell closure 
predicted by the HFB calculations and seen experimentally. This procedure can 
be applied only to nuclei for which the mass of the 20 
neighbouring nuclei is included in the mass table (see Eq.~\ref{eq_gk}).
We iterate 5 times, in order to achieve some degree of convergence. 
The resulting $S_{2n}/2$ mass surface for all isotope chains with model HFB-24 
is shown in the lower panel of Fig.~\ref{fig_A2}. The success of our procedure
becomes apparent on comparing with the upper panel of this same figure, which
shows the same quantities for the original HFB-24 model, before filtering.
It will be seen that there has been a significant reduction in the noise
level. 

Fig.~\ref{fig_A3} shows for all 8509 nuclei included in the HFB-24 mass 
table the actual shift in the calculated masses brought about by our
noise-filtering procedure. The rms value of the shift over all three models is 0.20~MeV, although
for specific cases (notably light nuclei) the shift can amount to more than
1 MeV. 

As for the impact of our procedure on the experimentally known nuclei,
the rms deviation for all 2353 known nuclei increases by no more than
4~keV with respect to the original HFB table, and for the 257 most
neutron-rich nuclei with $S_n\le 5$~MeV an improvement of 7~keV is found.
Not surprisingly, there is a slight improvement by some 20~keV in the rms deviations of the
differential quantities $S_n$ and $Q_\beta$.

We stress that we are not using the Garvey-Kelson method~\cite{ga69} in its 
traditional role as a means of extrapolating masses from known to unknown 
nuclei: to do so would have led to enormous deviations from the HFB results. 
Rather, this extrapolation is performed at the level of the original HFB calculations, 
once the forces have been 
obtained. In other words, the Garvey-Kelson method, in the form of 
Eq.~(\ref{eq_gk}), is applied to all the HFB-calculated masses in the same way,
regardless of whether they refer to measured or unmeasured nuclei. Thus the HFB
result for {\it every} nucleus leaves its imprint on the final noise-filtered
result.

\end{document}